\documentclass[twocolumn,showpacs,preprintnumbers,amsmath,amssymb]{revtex4}

\usepackage{graphicx}
\usepackage{dcolumn}
\usepackage{bm}
\usepackage{gensymb}

\begin{document}


\title{Efficient Non-Resonant Absorption in Thin Cylindrical Targets: Experimental Evidence for Longitudinal Geometry}

\author{A. Akhmeteli}
\email{akhmeteli@ltasolid.com}
 \homepage{http://www.akhmeteli.org}
\affiliation{%
LTASolid Inc.\\
10616 Meadowglen Ln 2708, Houston, TX 77042, USA
}%
\author{N. G. Kokodiy}%
\email{kokodiy.n.g@gmail.com}
\author{B.V. Safronov}%
\author{V.P. Balkashin}%
\author{I.A. Priz}%
\affiliation{%
Kharkov National University\\
Ukraine, Kharkov, Svobody sq., 4
}%
\author{A. Tarasevitch}%
\affiliation{%
University of Duisburg-Essen, Institute of Experimental Physics\\
Lotharstr. 1, 47048 Duisburg, Germany
}%

\date{\today}

\begin{abstract}
Experiments provide a qualitative confirmation of significant absorption of a wide electromagnetic beam propagating along a thin conducting cylinder (the diameter of the cylinder can be orders of magnitude less than the beam waist width). This new physical effect can be used for numerous applications, such as pumping of active media of short-wavelength lasers and creation of low-density channels in the lower atmosphere.
\end{abstract}

\maketitle

\section{\label{sec:level1}Introduction}

A theoretical possibility of non-resonant, fast, and efficient heating of very thin conducting cylindrical targets by coaxial broad electromagnetic beams was described in Ref.~\cite{Akhm1} (see also the sections on the "longitudinal geometry" in Refs.~\cite{Akhm10,Akhm111} and references there). The diameter of the cylinder can be orders of magnitude smaller than the beam waist of the electromagnetic radiation. Efficient heating takes place in several broad domains of parameters (see the exact conditions in Refs.~\cite{Akhm10,Akhm111}). This possibility can be used for such applications as pumping of active media of short-wavelength lasers and creation of low-density channels in the lower atmosphere (Ref.~\cite{Akhm1}). Recently, experimental confirmation of efficient heating of extremely thin conducting cylindrical targets by broad electromagnetic beams was obtained in Ref.~\cite{Akhm113} for the transverse geometry of Refs.~\cite{Akhm10,Akhm111}.

\begin{figure*}
\includegraphics{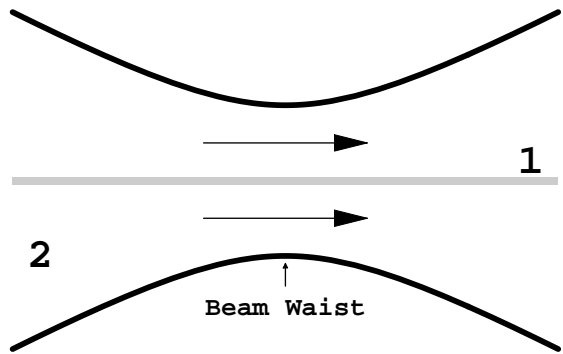}
\caption{\label{fig:gr1}The longitudinal geometry. Electromagnetic beam 2 propagates along the axis of target cylinder 1.}
\end{figure*}

The longitudinal geometry of target irradiation by an electromagnetic beam is illustrated in Fig.~\ref{fig:gr1}.
In this work we present a qualitative experimental confirmation of the predictions of Refs.~\cite{Akhm1,Akhm10,Akhm111}).

\section{\label{sec:level1-2}The experimental setup}

\begin{figure*}
\includegraphics[width=170mm]{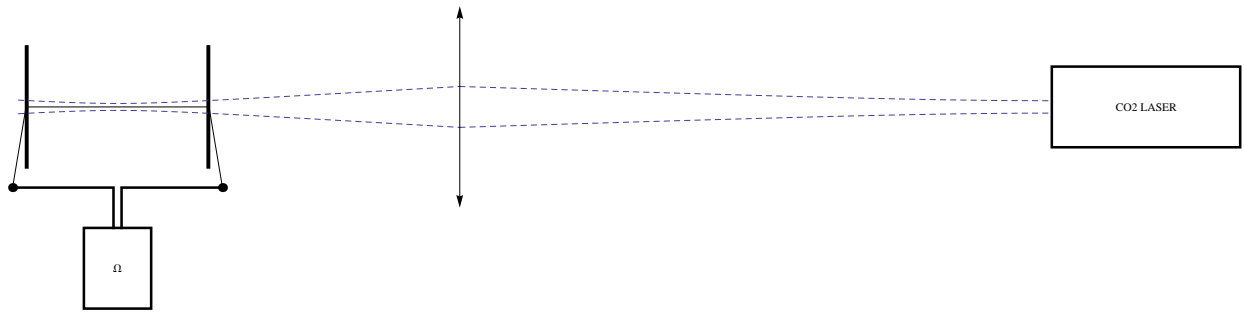}
\caption{\label{fig:picnew2}The experimental setup (not to scale).}
\end{figure*}

The experimental setup is shown schematically in Fig.~\ref{fig:picnew2}.

The CO$_{\textrm{2}}$ laser emits infrared radiation (the wavelength $\lambda$ is 10.6 micron). The thin wire was mounted using crosswires in vertical planes at each end. Each crosswire consisted of two mutually orthogonal thin nickel wires (length -- about 60 mm, diameter -- 50 micron). The wires were placed at $45\degree$  to the vertical. The horizontal wire was placed upon these crosswires and had to be pulled taut to ensure its stability in the course of heating. Previously, a different wire mounting method was used (see the previous version of this preprint). The parameters of the horizontal wire are given in Table 1.

\begin{center}
  \begin{tabular}{| c | c | c|c|c| }
    \hline
    Material & $D$, $\mu$m & $L$, m & $\alpha_r$, K$^{-1}$ & $\alpha_p$, W/(m$\cdot$K) \\ \hline
    Pt & 20 & 1.1 & $4\cdot 10^{-3}$ & $4\cdot 10^{-2}$  \\
    \hline
  \end{tabular}
\end{center}

Table 1. Wire material, diameter $D$, length $L$, temperature coefficient of resistance $\alpha_r$, linear heat exchange coefficient $\alpha_p$.

The linear heat exchange coefficient depends weakly on the material and the wire diameter in this range of parameters. It was measured through heating the wire with direct current.

The beam is focused by a ZnSe lens (the focal length $f$ is 1700 mm). The electrical resistance of the wire is measured with an ohmmeter. The parameters of the beam and the wire provide efficient absorption for the platinum wire  (case 2.1.2.1, Eq. (167) of Ref.~\cite{Akhm111}, version 1).

\section{\label{sec:level1-3}Measurement and computation methods}
The wire is heated by the radiation, and the initial electrical resistance of the wire $R_0$  changes by $\Delta R$. The average wire temperature increase  $\Delta T$ corresponding to  $\Delta R$ was calculated as
\begin{equation}
\Delta T=\frac{\Delta R}{\alpha_r R_0}.\label{eq:2}
\end{equation}

On the other hand, the steady  state wire temperature increase depends on the absorbed power $P_a$ and the conditions of heat exchange with the environment (the Newton's law of cooling) \cite{Inc}:
\begin{equation}
\Delta T=\frac{P_a}{a_p L},\label{eq:3}
\end{equation}
where $L$ is the length of the wire between the crosswires, $a_p$ is the linear heat exchange coefficient for the wires used in the experiment.

It follows from Eqs.~(\ref{eq:2},\ref{eq:3}) that
\begin{equation}
P_a=\frac{\alpha_p L}{\alpha_r}\frac{\Delta R}{R_0}.\label{eq:4}
\end{equation}
Therefore, the efficiency of absorption of laser beam power in the wire equals:
\begin{equation}
K=\frac{P_a}{P}=\frac{\alpha_p L}{\alpha_r P}\frac{\Delta R}{R_0},\label{eq:5}
\end{equation}
where P is the power in the laser beam.

\begin{figure*}
\includegraphics[width=170mm]{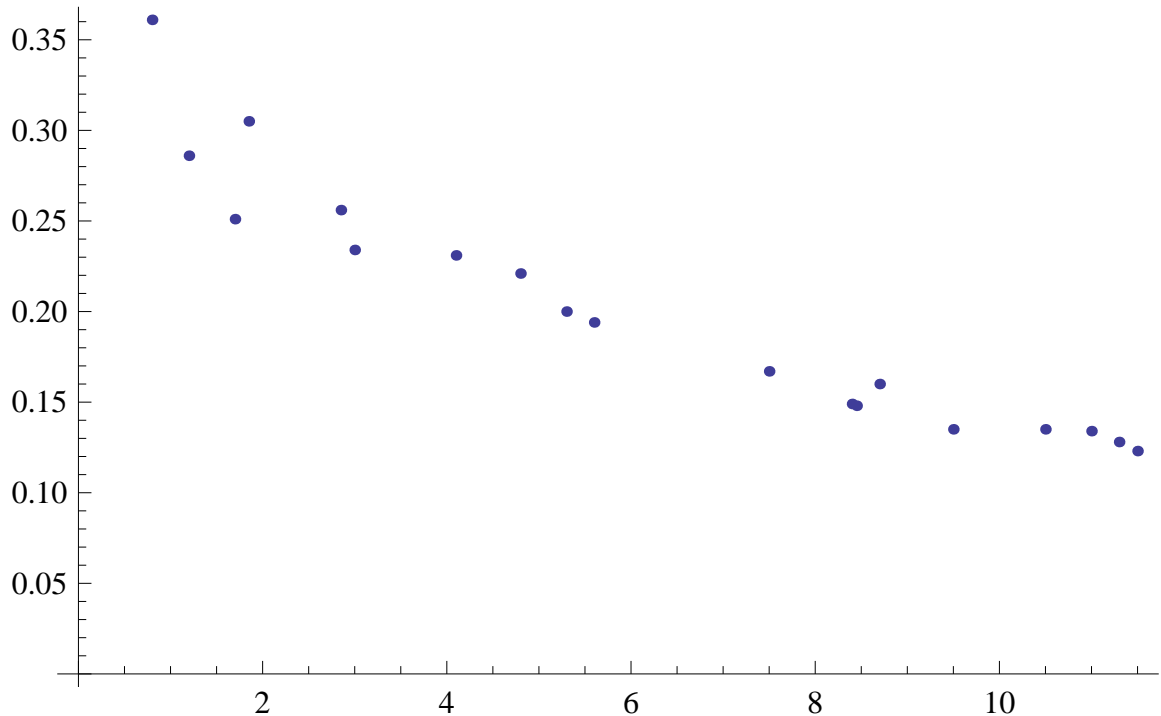}
\caption{\label{fig:processingextra}Absorption efficiency vs beam power, W.}
\end{figure*}

The results of the experiments are presented in Fig.~\ref{fig:processingextra}.

The measured absorption efficiency (13-36\%) was compared to the absorption efficiency of 58\% computed using the precise theoretical method for a gaussian beam of Refs.~\cite{Akhm1,Akhm10,Akhm111} (see the details at the end of this section). Thus, the experimental results are in qualitative agreement with the theoretical predictions derived for gaussian beams.  It should be emphasized that the efficiency achieved in the experiment is quite high, as the measured width of the beam's waist (1.2 mm at 1/e intensity level) is two orders of magnitude greater than the wire diameter (20 micron).

The exact cause of the nonlinearity (significant dependence of the absorption efficiency on the beam power) has not been established yet, but wire elongation due to heating is a candidate, as it could dramatically decrease absorption efficiency at higher beam power levels: while this elongation was small for the temperature increase in the experiment, it could lead to wire sagging by several millimeters. As a result, the wire could move out of the beam, at least partially. To avoid this, the wire was connected to a spring at one end to provide compensation of the elongation.

The incident gaussian beam was modeled by an exact solution of the Maxwell equations in free space (equations 176-178 of Ref.~\cite{Akhm111}, version 1). This solution describes a circularly polarized beam, but it was used for the linearly polarized beam of the laser in this preliminary computation, as the results depend on polarization very weakly in this geometry. The total power absorbed in the cylinder was computed using equations 186-190 of Ref.~\cite{Akhm111}, v. 1, and equations 99 and 101 of Ref.~\cite{Akhm10}, v. 1, but the integration in equation 186 of Ref.~\cite{Akhm111}, v. 1, was performed over the finite length of the wire, rather than from $-\infty$ to $\infty$.

\section{\label{sec:level1-3}Conclusion}

The results of the experiments demonstrate the feasibility of efficient heating of thin cylindrical targets with a coaxial electromagnetic beam with a waist width several orders of magnitude greater than the diameter of the target. To this end, there needs to be a match between the diameter of the target, its conductivity, and the wavelength. However, the conditions of efficient heating are non-resonant and therefore very promising for numerous applications. The heating efficiency of tens percent can be achieved for very thin targets.

\end{document}